\pdfoutput=1
\documentclass[preprint,preprintnumbers,amsmath,amssymb,floatfix,endfloats*]{revtex4}

\usepackage{graphicx}
\usepackage{dcolumn}
\usepackage{bm}
\usepackage{amsfonts}

\DeclareMathAlphabet{\mathsfsl}{OT1}{cmr}{bx}{it}
\begin{document}
%
\title{Fatigue failure of amorphous alloys under cyclic shear deformation}
\author{Nikolai V. Priezjev$^{1,2}$}
\affiliation{$^{1}$Department of Civil and Environmental
Engineering, Howard University, Washington, D.C. 20059}
\affiliation{$^{2}$Department of Mechanical and Materials
Engineering, Wright State University, Dayton, OH 45435}
\date{\today}
\begin{abstract}

The accumulation of plastic deformation and flow localization in
amorphous alloys under periodic shear are investigated using
molecular dynamics simulations. We study a well-annealed binary
mixture of one million atoms subjected to oscillatory shear
deformation with strain amplitudes slightly above a critical value.
We find that upon approaching a critical strain amplitude from
above, the number of shear cycles until the yielding transition is
well described by a power-law function. Remarkably, the potential
energy at the end of each cycle as a function of the normalized
number of cycles is nearly independent of the strain amplitude,
which allows for estimation of the fatigue lifetime at a given
strain amplitude. The analysis on nonaffine displacements of atoms
elucidates the process of strain localization, including
irreversible rearrangements of small clusters until the formation of
a system-spanning shear band.

\vskip 0.5in


\end{abstract}

\maketitle

\section{Introduction}

The prediction of stability and lifetime of amorphous alloys under
repeated stress or strain deformation is important for various
structural applications~\cite{Ashby06,Sha2020}. Although
multicomponent alloys like metallic glasses possess a number of
advantageous properties, such as high strength and large elastic
strain limit, their resistance to fatigue damage is relatively
poor~\cite{Egami13, Ritchie99, Menzel06, Tanaka22}. The failure
mechanism in metallic glasses involves the formation of nanoscale
shear bands where plastic strain becomes strongly localized, which
in turn might lead to propagation of microscale cracks~\cite{Liaw05,
Wang17, Hufnagel16, Shi19}. At the atomic level, the elementary
plastic deformation in amorphous solids consists of rapid
rearrangement of a small cluster of particles or shear
transformation~\cite{Spaepen77, Argon79}. Notably, the results of
numerical simulations of the fibre bundle model have shown that the
fatigue failure under repeated loading of heterogeneous materials
occurs after a number of cycles, and the fatigue lifetime has a
power-law dependence on the loading amplitude~\cite{Herrmann07}.
More recently, using two models of elastoplastic rheology, it was
demonstrated that cyclically sheared amorphous materials initially
accumulate low levels of damage in the form of spatial strain
heterogeneity, which is followed by a sudden catastrophic material
failure via shear band formation~\cite{Fielding22}. However, in
spite of the considerable modeling and experimental efforts, the
precise determination of the critical loading amplitude and fatigue
lifetime remains a challenging problem.

\vskip 0.05in

During the last decade, the effect of cyclic loading on the yielding
transition, structural relaxation, and flow localization in
amorphous materials was extensively studied using atomistic
simulations~\cite{Priezjev13, Reichhardt13, Sastry13, IdoNature15,
GaoNano15, Priezjev16, Kawasaki16, Priezjev16a, Sastry17,
Priezjev17, Priezjev18, Priezjev18a, Sastry19band, PriezSHALT19,
Priez20ba, KawBer20, NVP20altY,  Priez20delay, BhaSastry21,
Priez21var, PriezCMS21, Peng22, Procaccia22, PriezJNCS22, Maloney22,
Barrat23}. Interestingly, it was demonstrated that in the athermal
limit, amorphous solids under small-amplitude oscillatory shear
evolve into the so-called limit cycles, where trajectories of atoms
become exactly reversible after one or more periods, and the number
of cycles to reach periodic behavior diverges upon approaching a
critical strain amplitude from below~\cite{IdoNature15,
Reichhardt22}.  On the other hand, periodic deformation at strain
amplitudes above a critical value leads to yielding and flow
localization after a number of cycles~\cite{GaoNano15, Sastry17,
Priezjev17, Sastry19band, NVP20altY}. In general, the number of
cycles until the yielding transition depends on the degree of
annealing, temperature, system size, strain amplitude and frequency.
In particular, it was found that the number of cycles to failure
increases upon increasing frequency~\cite{GaoNano15} or glass
stability~\cite{Priez20ba, Priez20delay}, and by decreasing strain
amplitude towards a critical value~\cite{GaoNano15, Priezjev17}. In
addition, the number of fatigue cycles can be reduced by
periodically alternating shear orientation in two or three spatial
dimensions~\cite{NVP20altY} or by occasionally increasing strain
amplitude above a critical value~\cite{Priez21var}. Despite recent
progress, however, the processes of damage accumulation and
formation of shear bands during cyclic loading near a critical
strain amplitude remain not fully understood.

\vskip 0.05in

In this paper, the influence of repeated shear strain on plastic
deformation and yielding transition in a disordered solid is studied
via molecular dynamics (MD) simulations. We consider a well-annealed
binary glass subjected to oscillatory shear deformation at strain
amplitudes slightly above a critical value. It will be shown that
the number of shear cycles to reach the yielding transition
increases approximately as a power-law function when the strain
amplitude approaches the critical value. Moreover, we find that the
potential energy at zero strain for different strain amplitudes is
well described by a single function of the normalized number of
cycles. In turn, the appearance of local plastic events and the
formation of a shear band at the yielding transition are quantified
via the fraction of atoms with large nonaffine displacements during
one shear cycle.

\vskip 0.05in

The rest of this paper is organized as follows. The details of
molecular dynamics simulations as well as the oscillatory shear
deformation protocol are described in the next section. The analysis
of shear stress, potential energy, and nonaffine displacements is
presented in section\,\ref{sec:Results}. A brief summary is provided
in the last section.

\section{Molecular dynamics simulations}
\label{sec:MD_Model}

In this study, the amorphous alloy was modeled via the standard
Kob-Andersen (KA) binary mixture composed of 80\,\% of atoms of type
A and 20\,\% of type B~\cite{KobAnd95}. The total number of atoms is
$10^6$. In this model, the interaction between atoms of types
$\alpha,\beta=A,B$ is defined via the Lennard-Jones (LJ) potential:
\begin{equation}
V_{\alpha\beta}(r)=4\,\varepsilon_{\alpha\beta}\,\Big[\Big(\frac{\sigma_{\alpha\beta}}{r}\Big)^{12}\!-
\Big(\frac{\sigma_{\alpha\beta}}{r}\Big)^{6}\,\Big],
\label{Eq:LJ_KA}
\end{equation}
where the parameters are set to $\varepsilon_{AA}=1.0$,
$\varepsilon_{AB}=1.5$, $\varepsilon_{BB}=0.5$, $\sigma_{AA}=1.0$,
$\sigma_{AB}=0.8$, $\sigma_{BB}=0.88$, and
$m_{A}=m_{B}$~\cite{KobAnd95}. A similar parametrization was used by
Weber and Stillinger to study structure and dynamics of the
amorphous metal-metalloid alloy
$\text{Ni}_{80}\text{P}_{20}$~\cite{Weber85}. All physical
quantities are reported in the units of length, mass, energy, and
time, as follows: $\sigma=\sigma_{AA}$, $m=m_{A}$,
$\varepsilon=\varepsilon_{AA}$, and
$\tau=\sigma\sqrt{m/\varepsilon}$. The MD simulations were carried
out using the LAMMPS parallel code with the integration time step
$\triangle t_{MD}=0.005\,\tau$ and the cutoff radius
$r_{c}=2.5\,\sigma$~\cite{Allen87,Lammps}.

\vskip 0.05in


The sample preparation procedure and the deformation protocol are
similar to the ones reported in the previous MD
study~\cite{Priez20ba}. More specifically, the binary mixture was
first placed in a cubic box of linear size $L=94.10\,\sigma$ and
equilibrated at the temperature $T_{LJ}=1.0\,\varepsilon/k_B$ and
density $\rho=\rho_A+\rho_B=1.2\,\sigma^{-3}$ using the
Nos\'{e}-Hoover thermostat and periodic boundary
conditions~\cite{Allen87,Lammps}. For reference, the critical
temperature of the KA model at the density $\rho=1.2\,\sigma^{-3}$
is $T_g=0.435\,\varepsilon/k_B$~\cite{KobAnd95}. Then, the sample
was cooled with computationally slow rate of
$10^{-5}\varepsilon/k_{B}\tau$ from $T_{LJ}=1.0\,\varepsilon/k_B$ to
$0.01\,\varepsilon/k_B$ at constant density $\rho=1.2\,\sigma^{-3}$.
Right after cooling, the glass was subjected to oscillatory shear
deformation along the $xz$ plane, as follows:
\begin{equation}
\gamma_{xz}(t)=\gamma_0\,\text{sin}(2\pi t/T ),
\label{Eq:shear}
\end{equation}
where $\gamma_0$ is the strain amplitude and $T=5000\,\tau$ is the
oscillation period. The simulations were performed for strain
amplitudes $0.069 \leqslant \gamma_0 \leqslant 0.075$ at
$T_{LJ}=0.01\,\varepsilon/k_B$ and $\rho=1.2\,\sigma^{-3}$. The
results for the potential energy, shear stress, and nonaffine
displacements of atoms are reported only for one realization of
disorder because of the considerable computational burden. As an
example, it took about 36 days to simulate 800 shear cycles at the
strain amplitude $\gamma_0=0.069$ using 400 processors in parallel.

\section{Results}
\label{sec:Results}


Recent studies have shown that model glasses prepared by thermal
annealing can yield after a certain number of cycles at strain
amplitudes that are smaller than the yielding strain during uniform
shear deformation~\cite{Sastry17,Priez20ba}. The precise value of
the critical strain amplitude is difficult to determine numerically
due to a large number of cycles needed to reach the yielding
transition. Within the range of about three thousand cycles, it was
found that rapidly quenched binary glasses under cyclic loading
yield at the critical strain amplitude $\gamma_0=0.067$, regardless
of whether shear is applied along a single plane or periodically
alternated in two or three spatial dimensions~\cite{NVP20altY}. In
the present study, we consider a relatively large system of one
million atoms and subject a well-annealed KA glass to oscillatory
shear deformation at strain amplitudes slightly above the critical
value.

\vskip 0.05in

\begin{figure}[t]
\includegraphics[width=12.0cm,angle=0]{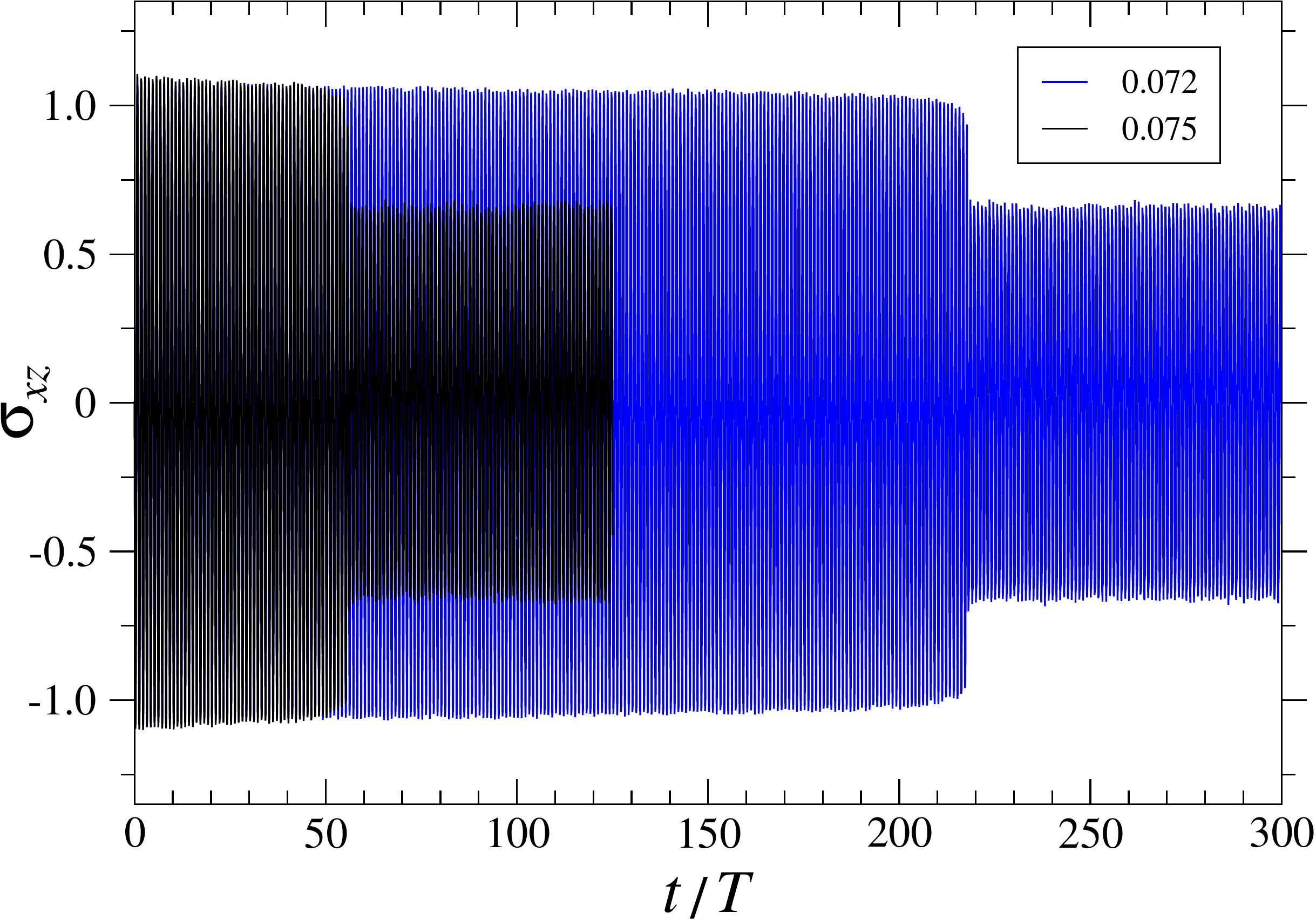}
\caption{(Color online) The shear stress, $\sigma_{xz}$ (in units
$\varepsilon\sigma^{-3}$), versus cycle number for strain amplitudes
$\gamma_0=0.072$ (blue curve) and $\gamma_0=0.075$ (black curve).
The period of oscillation is $T=5000\,\tau$. }
\label{fig:stress_xz_amp072_075}
\end{figure}

We first report the variation of shear stress along the $xz$ plane
as a function of time in Fig.\,\ref{fig:stress_xz_amp072_075} for
two values of the strain amplitude, i.e., $\gamma_0=0.072$ and
$0.075$.  It can be seen that in both cases, the amplitude of shear
stress oscillations slightly decreases upon continued loading until
a sudden drop during one shear cycle. Notice that the number of
cycles until yielding becomes greater upon decreasing strain
amplitude. Specifically, the yielding transition occurs during
218-th cycle for $\gamma_0=0.072$ and during 56-th cycle for
$\gamma_0=0.075$. By contrast, after the yielding transition, the
maximum shear stress is determined by the plastic flow within a
shear band. These results are consistent with those previously
reported for a well-annealed binary glass that was periodically
deformed for only 40 shear cycles at larger strain
amplitudes~\cite{Priezjev17}.

\vskip 0.05in

\begin{figure}[t]
\includegraphics[width=12.0cm,angle=0]{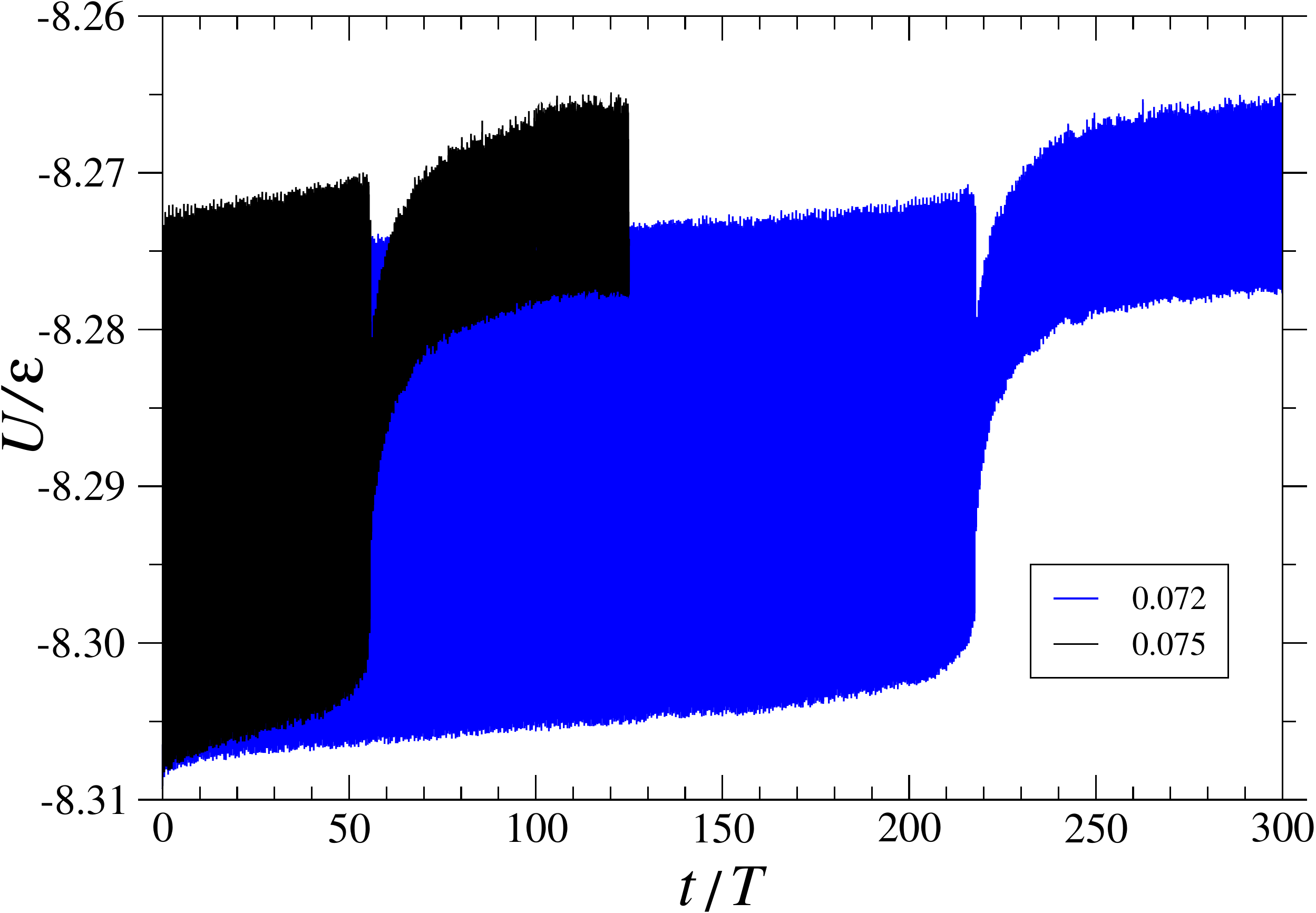}
\caption{(Color online) The potential energy per atom,
$U/\varepsilon$, as a function of time for strain amplitudes
$\gamma_0=0.072$ (blue curve) and $\gamma_0=0.075$ (black curve).
The period of oscillation is $T=5000\,\tau$. The binary glass was
initially prepared by cooling from $T_{LJ}=1.0\,\varepsilon/k_B$ to
$0.01\,\varepsilon/k_B$ with the rate of
$10^{-5}\varepsilon/k_{B}\tau$ at $\rho=1.2\,\sigma^{-3}$. }
\label{fig:poten_rem5_amp072_075}
\end{figure}

Along with shear stress, we plot in
Fig.\,\ref{fig:poten_rem5_amp072_075} the time dependence of the
potential energy for the same strain amplitudes, $\gamma_0=0.072$
and $0.075$, as in Fig.\,\ref{fig:stress_xz_amp072_075}. It is
evident that the yielding transition is associated with an abrupt
increase of the potential energy due to the formation of a shear
band across the system. It should be emphasized that for each strain
amplitude, the sudden change in shear stress and potential energy
occur at the same cycle number. One can further realize that before
yielding, the cyclic shear deformation results in a slow
accumulation of plastic events. This is reflected in a gradual
increase of the potential energy minima when strain is zero as a
function of the cycle number.

\vskip 0.05in

\begin{figure}[t]
\includegraphics[width=12.0cm,angle=0]{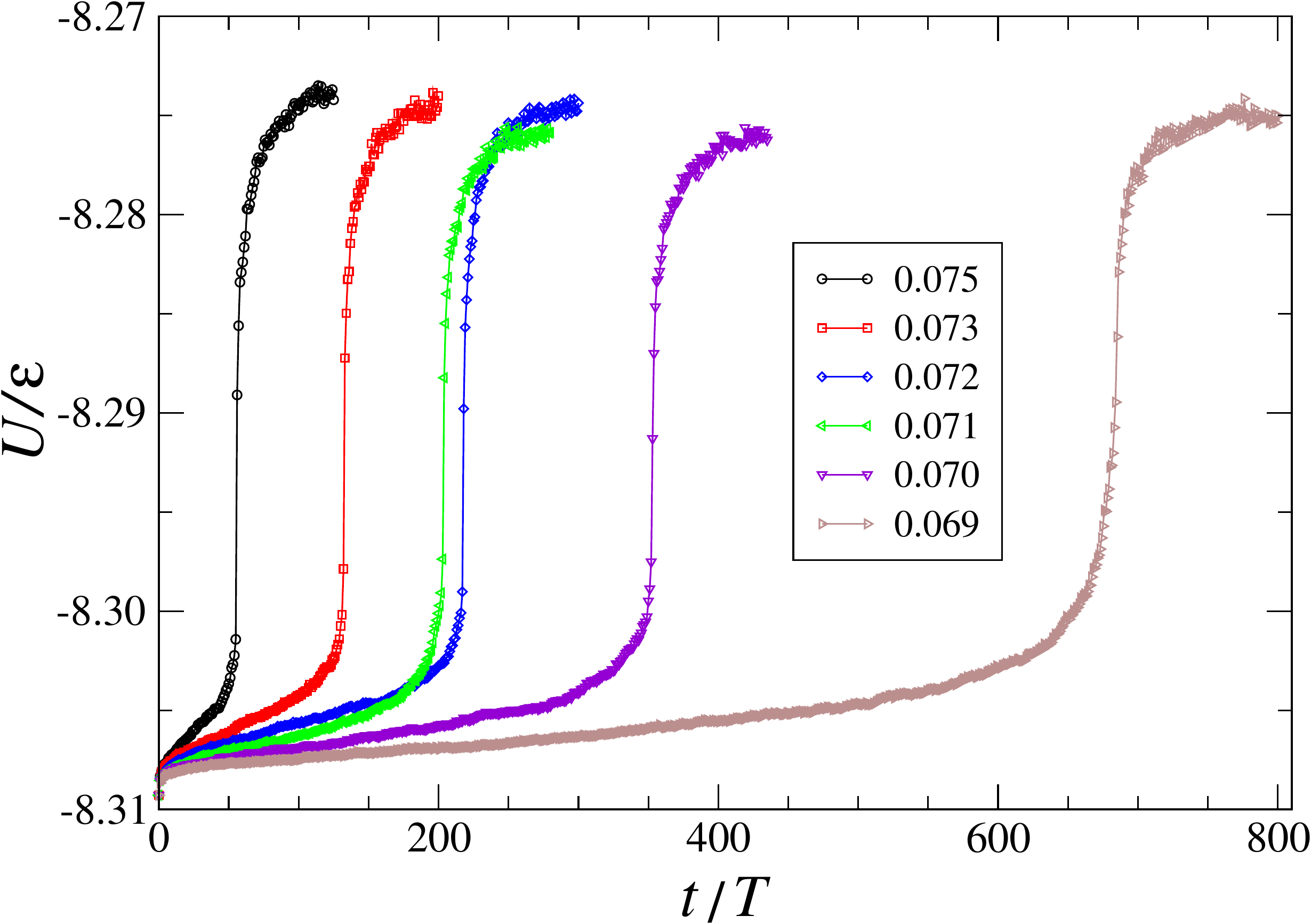}
\caption{(Color online) The potential energy minima at the end of
each shear cycle for strain amplitudes $\gamma_0=0.069$, $0.070$,
$0.071$, $0.072$, $0.073$, and $0.075$ (from right to left). The
oscillation period is $T=5000\,\tau$. }
\label{fig:poten_rem5_amps}
\end{figure}

Next, the potential energy minima at the end of each cycle are
presented in Fig.\,\ref{fig:poten_rem5_amps} for strain amplitudes
in the range $0.069 \leqslant \gamma_0 \leqslant 0.075$. Note that
the data at zero strain for $\gamma_0=0.072$ and $0.075$ are the
same as in Fig.\,\ref{fig:poten_rem5_amp072_075}. It can be clearly
observed in Fig.\,\ref{fig:poten_rem5_amps} that upon reducing
strain amplitude towards a critical value, the yielding transition
becomes significantly delayed. The exception to this trend is the
case of loading at the strain amplitude $\gamma_0=0.071$, where the
number of cycles until yielding is smaller than for
$\gamma_0=0.072$. In turn, the maximum number of cycles until the
yielding transition is $n_Y=685$ for the strain amplitude
$\gamma_0=0.069$. We comment that simulations at smaller strain
amplitudes, $\gamma_0 < 0.069$, were not carried out due to the high
computational cost.

\vskip 0.05in

\begin{figure}[t]
\includegraphics[width=12.0cm,angle=0]{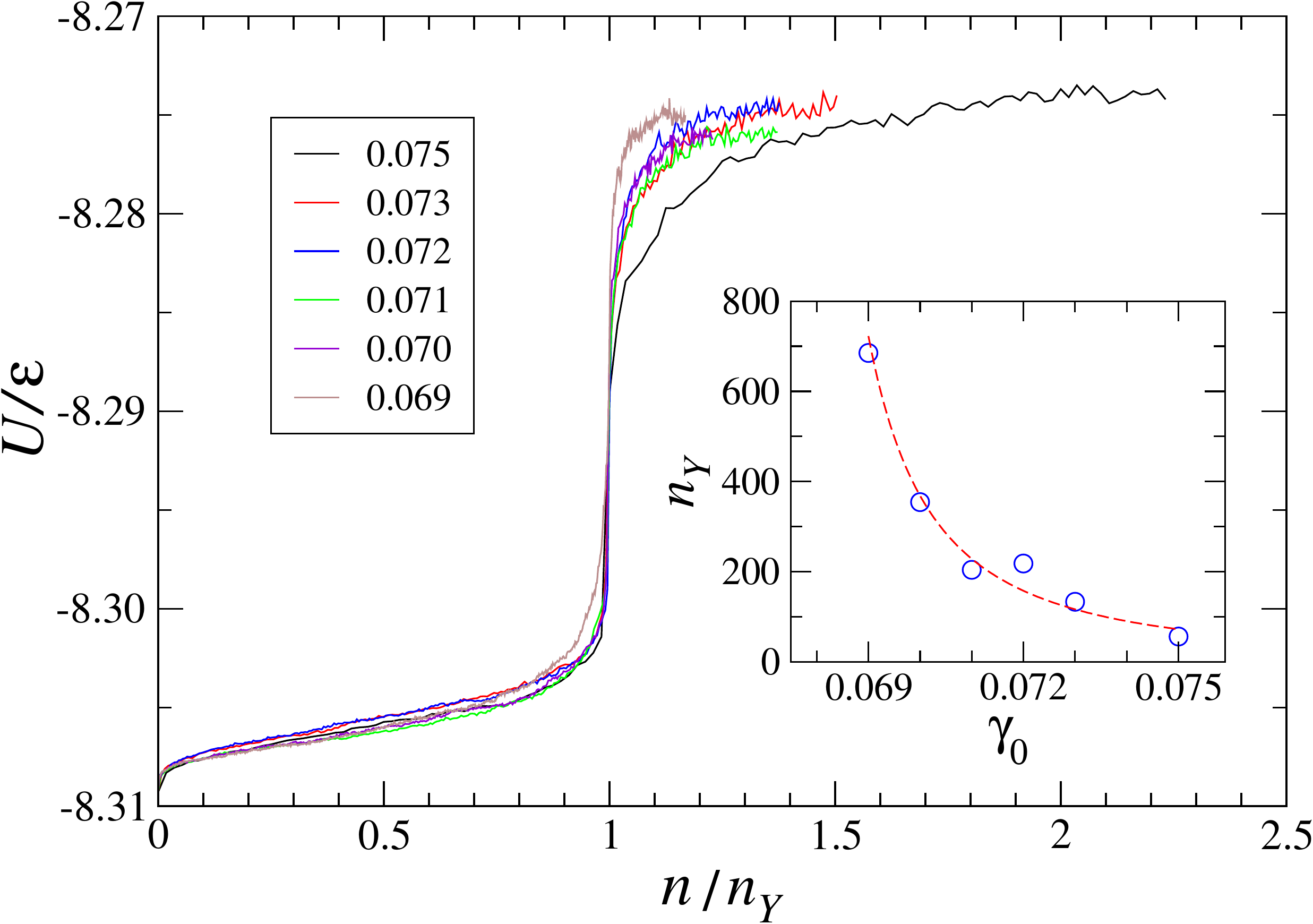}
\caption{(Color online) The potential energy minima as a function of
the ratio, $n/n_Y$, where $n=t/T$ is the cycle number and $n_Y$ is
the number of cycles to yield at a given strain amplitude. The same
data as in Fig.\,\ref{fig:poten_rem5_amps}. The symbols are omitted
for clarity. The inset shows $n_Y$ versus the strain amplitude
$\gamma_0$. The dashed red curve is the best fit to the data given
by Eq.\,(\ref{Eq:fit_nY}). }
%
\label{fig:poten_rem5_amps_scaled}
\end{figure}

The similarity of the functional form for potential energy minima
shown in Fig.\,\ref{fig:poten_rem5_amps} suggests a possibility of
rescaling the $\hat{x}$-coordinate by the number of cycles, $n_Y$,
required for the system to reach the yielding transition at a given
strain amplitude. Fig.\,\ref{fig:poten_rem5_amps_scaled} shows the
same potential energy curves as a function of the ratio $n/n_Y$.
Remarkably, the data for different $\gamma_0$ nearly collapse onto a
single curve when $n < n_Y$. The master curve is approximately
linear in the range $0.2 \lesssim n/n_Y \lesssim 0.8$, followed by a
steep increase due to accumulation of plastic events within a narrow
region that ultimately leads to flow localization when $n = n_Y$. On
the other end, the initial slope of the curve is determined by
irreversible rearrangements of group of atoms that settled at
relatively shallow energy minima after thermal annealing. In
practice, the function $U(n/n_Y)$ can be used to estimate $n_Y$ for
a binary glass loaded for a number of cycles ($n < n_Y$) at a strain
amplitude in the vicinity of the critical value.

\vskip 0.05in

Furthermore, the variation of $n_Y$ versus $\gamma_0$ is shown in
the inset of Fig.\,\ref{fig:poten_rem5_amps_scaled}. It is readily
apparent that the number of cycles until yielding increases
significantly when the strain amplitude approaches a critical value
from above. Moreover, the MD data are well described by the
power-law function, as follows:
\begin{equation}
n_Y=0.024\cdot(\gamma_0-0.067)^{-1.66},
\label{Eq:fit_nY}
\end{equation}
where the critical strain amplitude is taken to be $0.067$. This
value was determined previously for a smaller system of $60\,000$
atoms at $T_{LJ}=0.01\,\varepsilon/k_B$ and
$\rho=1.2\,\sigma^{-3}$~\cite{NVP20altY}. These results imply that
the number of cycles to reach the yielding transition might further
increase at lower strain amplitudes and possibly diverge in the case
of athermal systems~\cite{Sollich22}.

\vskip 0.05in


The local plastic events in disordered solids can be accurately
identified via the analysis of nonaffine displacements of
atoms~\cite{Falk98}. As a reminder, the nonaffine measure for
displacement of the $i$-th atom from $\mathbf{r}_{i}(t)$ to
$\mathbf{r}_{i}(t+\Delta t)$ is defined via the matrix
$\mathbf{J}_i$ that transforms positions of its neighboring atoms
and minimizes the following expression:
\begin{equation}
D^2(t, \Delta t)=\frac{1}{N_i}\sum_{j=1}^{N_i}\Big\{
\mathbf{r}_{j}(t+\Delta t)-\mathbf{r}_{i}(t+\Delta t)-\mathbf{J}_i
\big[ \mathbf{r}_{j}(t) - \mathbf{r}_{i}(t)  \big] \Big\}^2,
\label{Eq:D2min}
\end{equation}
where the summation is performed over $N_i$ atoms that are initially
located within $1.5\,\sigma$ from $\mathbf{r}_{i}(t)$. It should be
noted that the plastic rearrangement of neighboring atoms during the
time interval $\Delta t$ typically corresponds to values of the
nonaffine measure $D^2(t, \Delta t)$ greater than the cage size,
which is about $0.1\,\sigma$ for the KA binary glass at
$\rho=1.2\,\sigma^{-3}$~\cite{KobAnd95}.

\vskip 0.05in

\begin{figure}[t]
\includegraphics[width=12.0cm,angle=0]{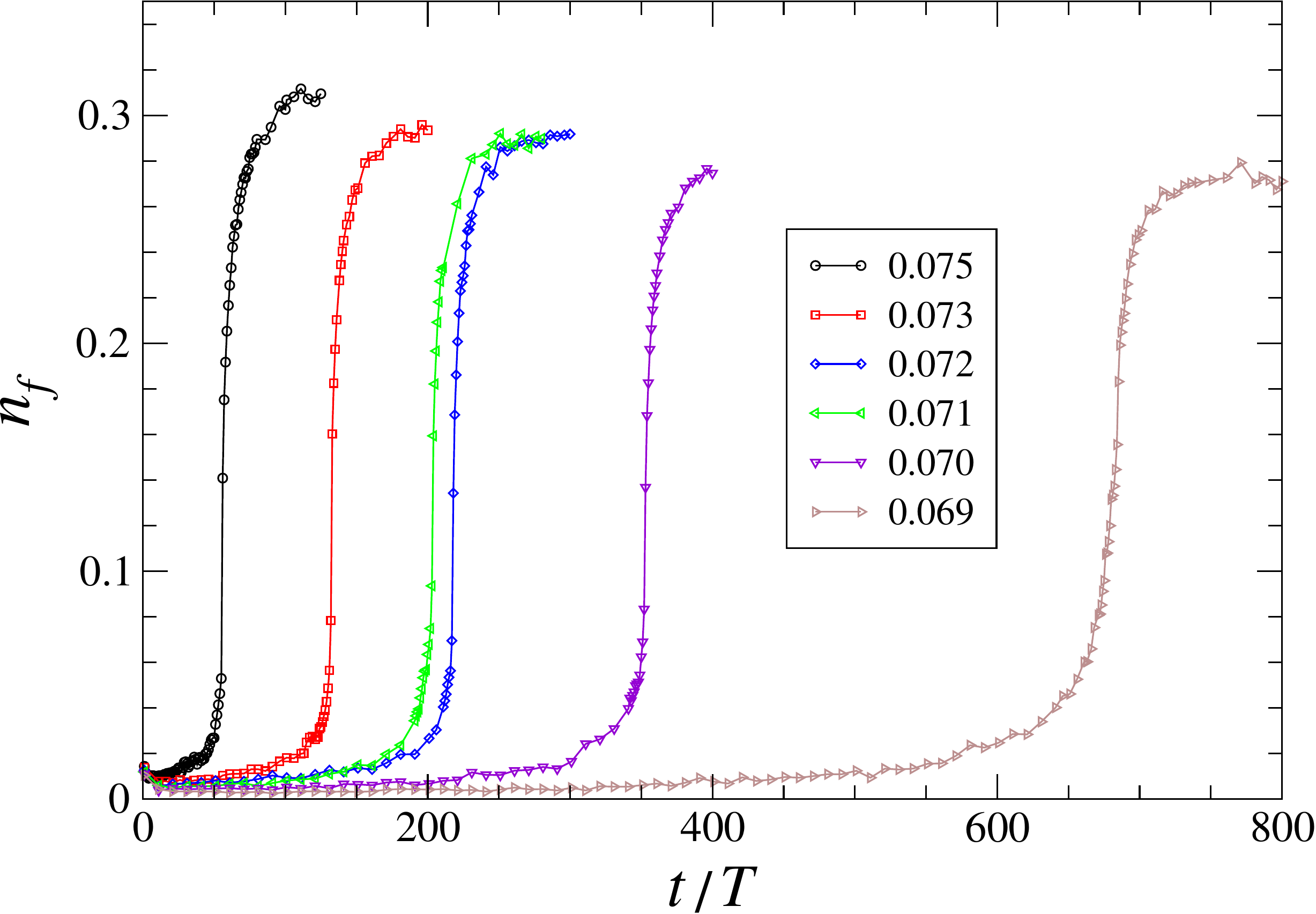}
\caption{(Color online) The fraction of atoms with large nonaffine
displacements, $D^2[(n-1)\,T,T]>0.04\,\sigma^2$, as a function of
the cycle number for the indicated values of the strain amplitude
$\gamma_0$. Here, $n=t/T$ is the cycle number, and $T=5000\,\tau$ is
the oscillation period. }
\label{fig:d2min_gt004_ncyc_amp}
\end{figure}

In Fig.\,\ref{fig:d2min_gt004_ncyc_amp} we show the fraction of
atoms with relatively large nonaffine displacements during one
cycle, $D^2[(n-1)T,T]>0.04\,\sigma^2$, for the strain amplitudes
$0.069 \leqslant \gamma_0 \leqslant 0.075$. We comment that the
nonaffine measure was evaluated only for selected cycles due to
excessive computational cost for the large system. It is clearly
observed that the shape of $n_f$ is similar to the dependence of
energy minima on the number of cycles shown in
Fig.\,\ref{fig:poten_rem5_amps}. Notice a small peak in $n_f$ during
the first cycle due to a number of atoms that become arranged in
shallow energy minima upon thermal annealing, and, as a result,
these atoms are prone to plastic rearrangement under shear
deformation. As expected, the yielding transition is clearly marked
by a sharp increase in the fraction $n_f$, indicating extended
plastic flow.

\vskip 0.05in

\begin{figure}[t]
\includegraphics[width=12.0cm,angle=0]{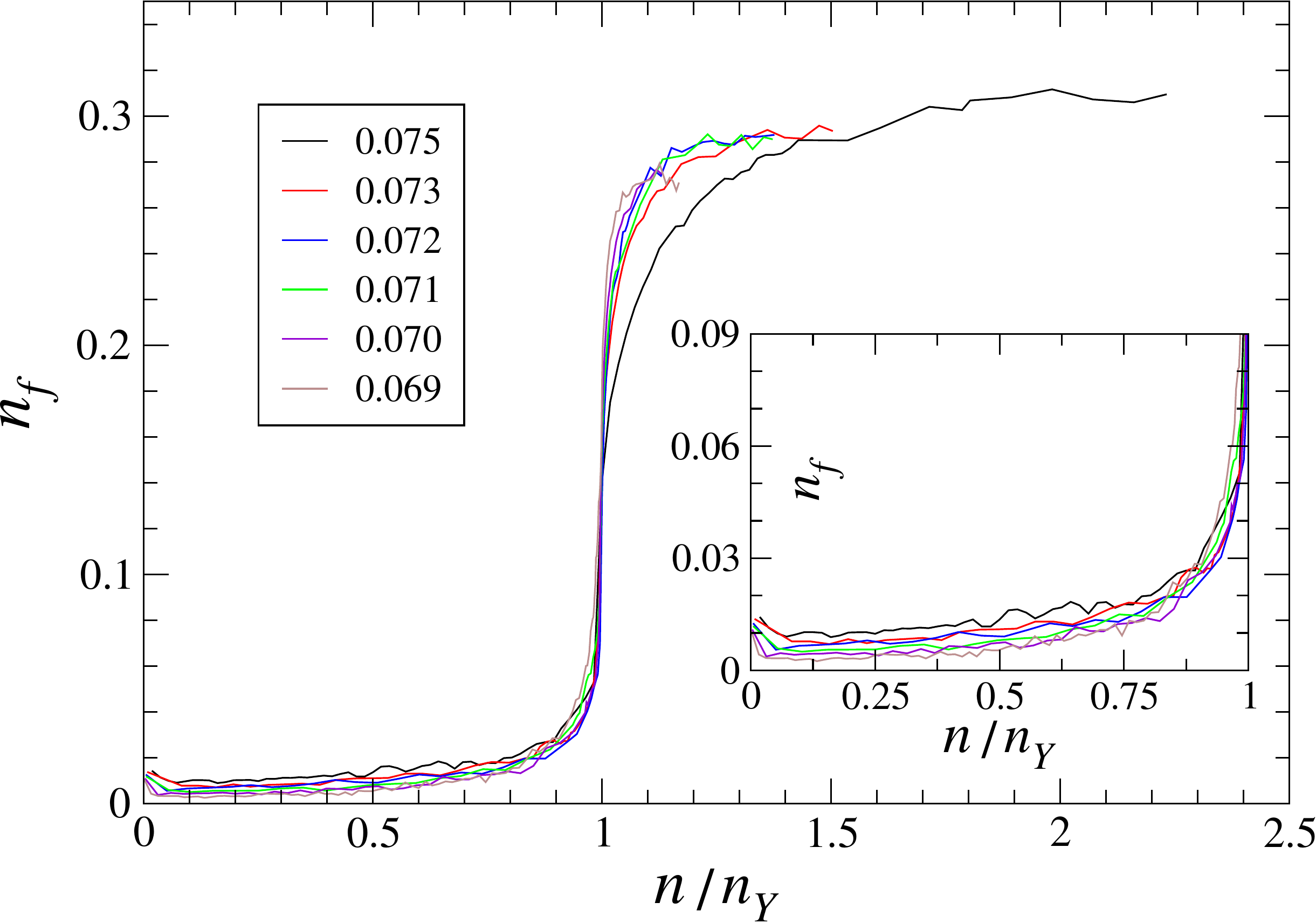}
\caption{(Color online) The fraction of atoms with
$D^2[(n-1)\,T,T]>0.04\,\sigma^2$ versus the ratio $n/n_Y$, where $n$
is the cycle number and $n_Y$ is the number of cycles until
yielding. The values of $n_Y$ are reported in
Fig.\,\ref{fig:poten_rem5_amps_scaled}. The symbols are not shown
for clarity. The inset shows an enlarged view of the same data for
$n/n_Y\leqslant1$. }
\label{fig:d2min_gt004_ncyc_amp_scaled}
\end{figure}

In analogy with the potential energy minima shown in
Fig.\,\ref{fig:poten_rem5_amps_scaled}, we replot the same data for
$n_f$ as a function of the ratio $n/n_Y$ in
Fig.\,\ref{fig:d2min_gt004_ncyc_amp_scaled}. It is evident that
fractions $n_f(n/n_Y)$ for different values of the strain amplitude
approximately follow a common curve. These results indicate that
plastic rearrangements of only about 1\,\% of atoms during the first
$n_Y/2$ cycles result in the increase of the potential energy
reported in Fig.\,\ref{fig:poten_rem5_amps_scaled}. A shear band
forms when $n_f \approx 0.14$ at $n=n_Y$. We also note that both
$n_f$ and $U$ increase and level out for $n
> n_Y$, which reflects widening of a shear band under cyclic shear.
In addition, a closer inspection of the data in the inset to
Fig.\,\ref{fig:d2min_gt004_ncyc_amp_scaled} reveals that, on
average, the fraction $n_f$ is slightly larger for cyclic loading at
higher strain amplitudes.

\vskip 0.05in

\begin{figure}[t]
\includegraphics[width=12.0cm,angle=0]{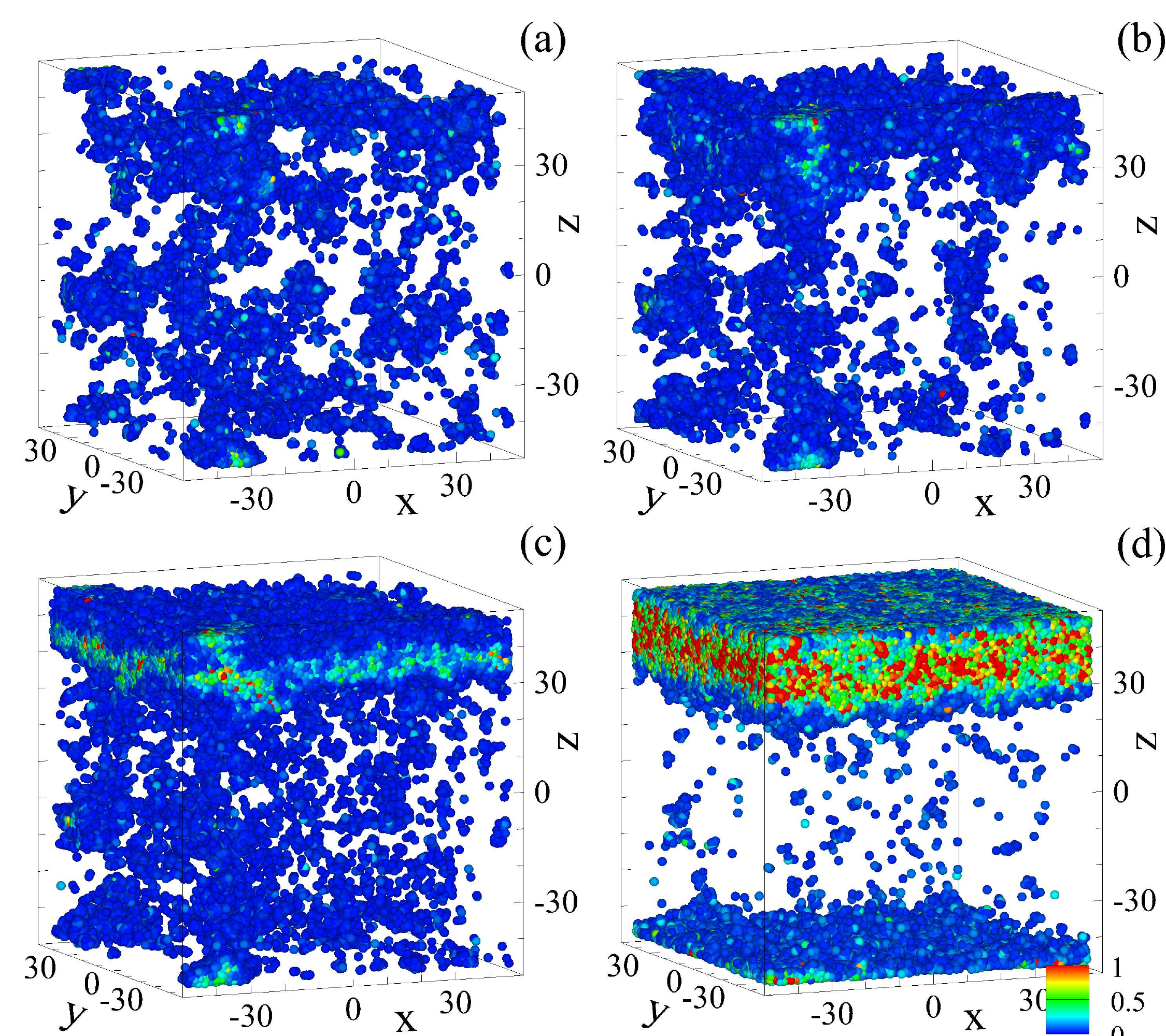}
\caption{(Color online) The atomic configurations of the binary
glass loaded at the strain amplitude $\gamma_0=0.072$. The nonaffine
displacements are shown for atoms with the nonaffine measure (a)
$D^2(200\,T,T)>0.04\,\sigma^2$, (b) $D^2(216\,T,T)>0.04\,\sigma^2$,
(c) $D^2(217\,T,T)>0.04\,\sigma^2$, and (d)
$D^2(224\,T,T)>0.04\,\sigma^2$. The legend color denotes the
magnitude of $D^2$. The atoms are not drawn to scale. }
\label{fig:snapshot_amp072}
\end{figure}
\begin{figure}[t]
\includegraphics[width=12.0cm,angle=0]{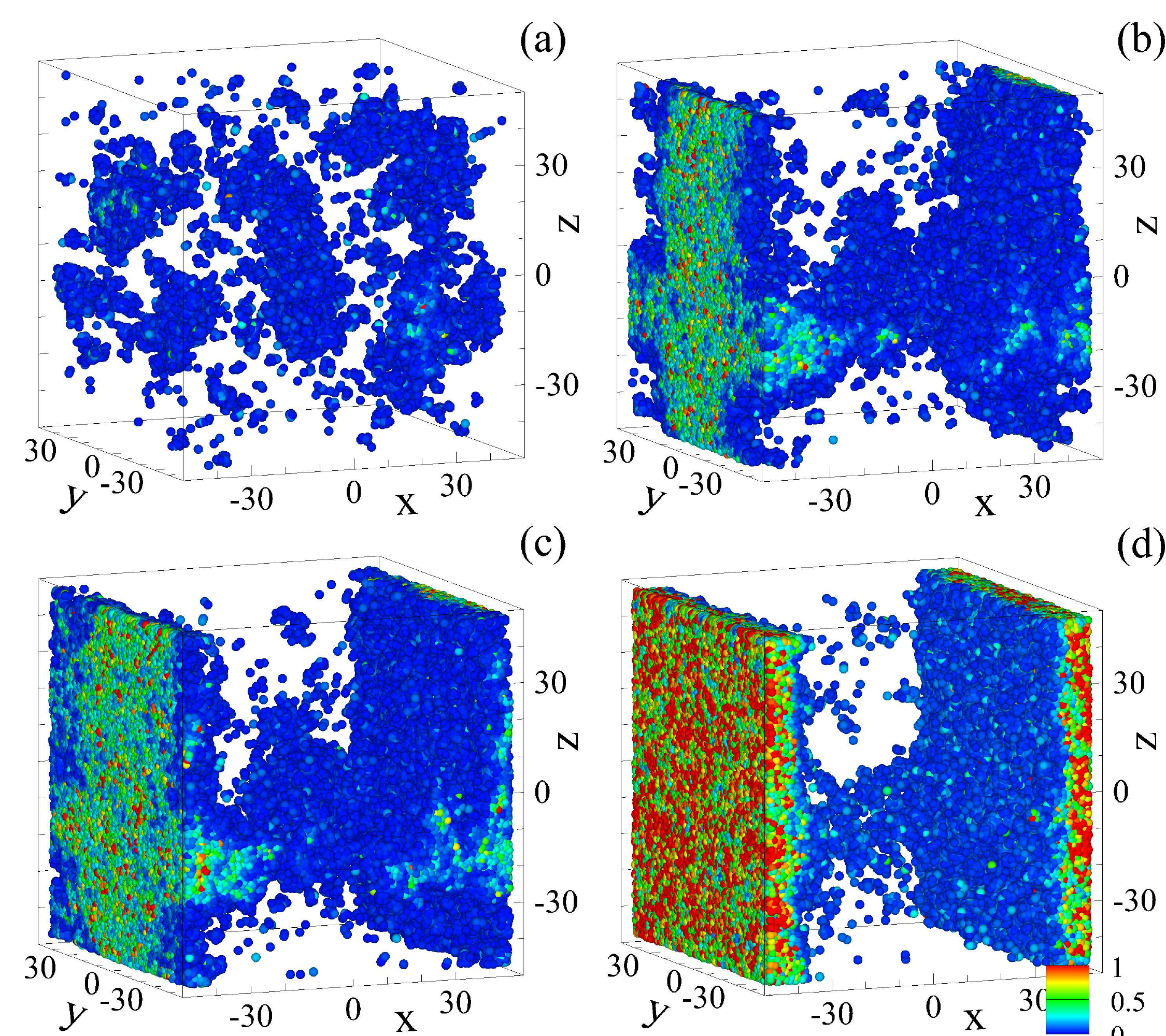}
\caption{(Color online) Snapshots of the well-annealed glass
subjected to cyclic shear with the strain amplitude
$\gamma_0=0.069$. The nonaffine measure is (a)
$D^2(600\,T,T)>0.04\,\sigma^2$, (b) $D^2(683\,T,T)>0.04\,\sigma^2$,
(c) $D^2(684\,T,T)>0.04\,\sigma^2$, and (d)
$D^2(689\,T,T)>0.04\,\sigma^2$. The colorcode indicates values of
$D^2$. }
\label{fig:snapshot_amp069}
\end{figure}

The spatial distribution of plastic rearrangements can be visualized
by plotting positions of atoms with large nonaffine displacements
during one shear cycle, i.e., $\Delta t = T$ in
Eq.\,(\ref{Eq:D2min}). For example, atomic configurations for
selected number of cycles are presented in
Fig.\,\ref{fig:snapshot_amp072} for the strain amplitude
$\gamma_0=0.072$ and in Fig.\,\ref{fig:snapshot_amp069} for
$\gamma_0=0.069$.  It can be seen in
Fig.\,\ref{fig:snapshot_amp072}\,(a) that before yielding, atoms
with $D^2(200\,T,T)>0.04\,\sigma^2$ are organized into small
clusters that are homogeneously distributed. Upon further loading,
the glass yields and a shear band forms along the $xy$ plane during
the 218-th cycle, as shown in Fig.\,\ref{fig:snapshot_amp072}\,(c).
During the next 7 cycles, the shear band becomes wider, which is
consistent with the increase in $n_f$ and $U$ after yielding
reported in
Figs.\,\ref{fig:poten_rem5_amps}--\ref{fig:d2min_gt004_ncyc_amp_scaled}.
Similar trends can be observed in Fig.\,\ref{fig:snapshot_amp069}
for cyclic loading at $\gamma_0=0.069$, except that the orientation
of the shear band is along the $yz$ plane. Also, the sequence of
snapshots for the strain amplitude $\gamma_0=0.075$ during the first
100 cycles was reported in the previous study~\cite{Priez20ba}.
Overall, the visualization of plastic events confirm our earlier
conclusions regarding the appearance of small clusters of atoms that
rearrange irreversibly after a full cycle, followed by the formation
of a shear band at the yielding transition, and its subsequent
widening upon continued loading.

\section{Conclusions}

In summary, the effect of oscillatory shear on the damage
accumulation and yielding transition was investigated using
molecular dynamics simulations. The binary glass was prepared by
cooling with a computationally slow rate deep into the glass phase
and then subjected to periodic shear deformation with strain
amplitudes slightly greater than a critical value. It was found that
the number of shear cycles until the yielding transition increases
approximately as a power-law function of the difference between the
strain amplitude and the critical value. We showed that the fatigue
process proceeds via a sequence of irreversible rearrangements of
small clusters of atoms until a sudden formation of a shear band at
the yielding transition. This behavior is reflected in the gradual
increase of the potential energy at the end of each cycle and a
steep increase near the yielding point. Furthermore, the potential
energy minima for different strain amplitudes closely follow a
master curve when plotted versus the normalized number of cycles.
The master curve can be used to estimate the fatigue lifetime for a
binary glass periodically deformed for only a small number of cycles
at a strain amplitude near the critical value.

\section*{Acknowledgments}

Financial support from the National Science Foundation (CNS-1531923)
is gratefully acknowledged. Molecular dynamics simulations were
carried out at Wright State University's Computing Facility and the
Ohio Supercomputer Center using the LAMMPS code~\cite{Lammps}.



%

%

%

%

%

%

%

%

\bibliographystyle{prsty}

\end{document}